%% file: main.tex
\documentclass[letterpaper]{article} 
\usepackage[]{aaai25}  
\usepackage{times}  
\usepackage{helvet}  
\usepackage{courier}  
\usepackage[hyphens]{url}  
\usepackage{graphicx} 
\usepackage{natbib}  
\usepackage{caption} 
\usepackage{algorithm}
\usepackage{algorithmic}
\usepackage{color}
\usepackage{tabularx}
\usepackage{booktabs}
\usepackage{xcolor}
\usepackage{times}
\usepackage{helvet}
\usepackage{courier}
\usepackage{graphicx}
\usepackage{amsmath}
\usepackage{amssymb}
\usepackage{cite}
\usepackage{booktabs}
\usepackage{multirow}
\usepackage{enumitem}

\usepackage{newfloat}
\usepackage{listings}
\DeclareCaptionStyle{ruled}{labelfont=normalfont,labelsep=colon,strut=off} 
\floatstyle{ruled}
\newfloat{listing}{tb}{lst}{}
\floatname{listing}{Listing}
\newcommand{\myname}[1]{\texttt{Pistis-RAG}}

\nocopyright

\title{Pistis-RAG: Enhancing Retrieval-Augmented Generation with Human Feedback}

\usepackage{bibentry}

\begin{document}
\author {
    Yu Bai,\textsuperscript{\rm 1}
    Yukai Miao,\textsuperscript{\rm 1}
    Li Chen,\textsuperscript{\rm 1}
    Dawei Wang,\textsuperscript{\rm 1}
    Dan Li,\textsuperscript{\rm 1,2}
    Yanyu Ren,\textsuperscript{\rm 2}
    \\ Hongtao Xie,\textsuperscript{\rm 3}
    Ce Yang,\textsuperscript{\rm 3}
    Xuhui Cai\textsuperscript{\rm 3}
}
\affiliations {
    \textsuperscript{\rm 1}Zhongguancun Laboratory, \\
    \textsuperscript{\rm 2}Tsinghua University, \\
    \textsuperscript{\rm 3}China Mobile Communications Group Co., Ltd. \\
    \{baiyu, miaoyk, lichen, wangdw\}@zgclab.edu.cn, tolidan@tsinghua.edu.cn, ryy23@mails.tsinghua.edu.cn, \\
    \{xiehongtao, yangce, caixuhui\}@chinamobile.com
}

\maketitle

\input{abstract}
\input{intro}

\input{related}

\input{approach}

\input{evaluation}
\input{discussion}

\input{conclusion}

\bigskip

\bibliography{aaai25}
\clearpage

\end{document}

%% file: abstract.tex
\begin{abstract}
RAG systems face limitations when semantic relevance alone does not guarantee improved generation quality. This issue becomes particularly evident due to LLMs' sensitivity \cite{lu2021fantastically} to the ordering of few-shot prompts, which can affect the model's performance. To address this challenge, aligning LLM outputs with human preferences using structured feedback—such as options to copy, regenerate, or dislike—offers a promising method for improvement. This feedback is applied to the entire list of inputs, rather than giving specific ratings for individual documents, making it a ``Listwide Labels'' Learning-to-Rank task. To address this task, we propose \myname{}, a new RAG framework designed with a content-centric approach to better align LLMs with human preferences. \myname{} effectively utilizes human feedback, enhancing content ranking and generation quality. To validate our framework, we use public datasets to simulate human feedback. This approach allows us to evaluate and refine our method effectively. Experimental results indicate that \myname{} improves alignment with human preferences relative to the baseline RAG system, with an increase of 6.06\% in MMLU (English) and 7.08\% in C-EVAL (Chinese) accuracy metrics. These results highlight \myname{}'s effectiveness in overcoming the limitations associated with traditional RAG approaches.
\end{abstract}

%% file: intro.tex
\section{Introduction}

Modern GenAI systems, such as ChatGPT, use long-term memory mechanisms to store critical question-and-answer data from interactions \cite{openai_memory_controls}. This stored information is utilized to generate few-shot prompts through the RAG process. By integrating retrieval-based information, RAG improves the effectiveness of these prompts, allowing LLMs to produce more relevant and accurate responses with minimal initial input \cite{lewis2020retrieval, borgeaud2022improving}.

Current research on RAG primarily focuses on enhancing the efficiency of few-shot prompting through improvements to the retriever component \cite{chen2024bge, glass2022re2g, kandpal2023large}. However, research addressing the alignment of RAG systems with human values remains limited. Since the primary objective of GenAI systems is to fulfill human needs, aligning these systems with human values is essential.

Extensive research has been conducted on aligning LLMs with human behavior using methodologies such as Reinforcement Learning with Human Feedback (RLHF) \cite{ouyang2022training}. Further investigation is necessary to explore how RAG systems can be more effectively aligned with human values and address related ethical considerations.

\begin{figure}[htbp]
\centering
\includegraphics[width=\columnwidth]{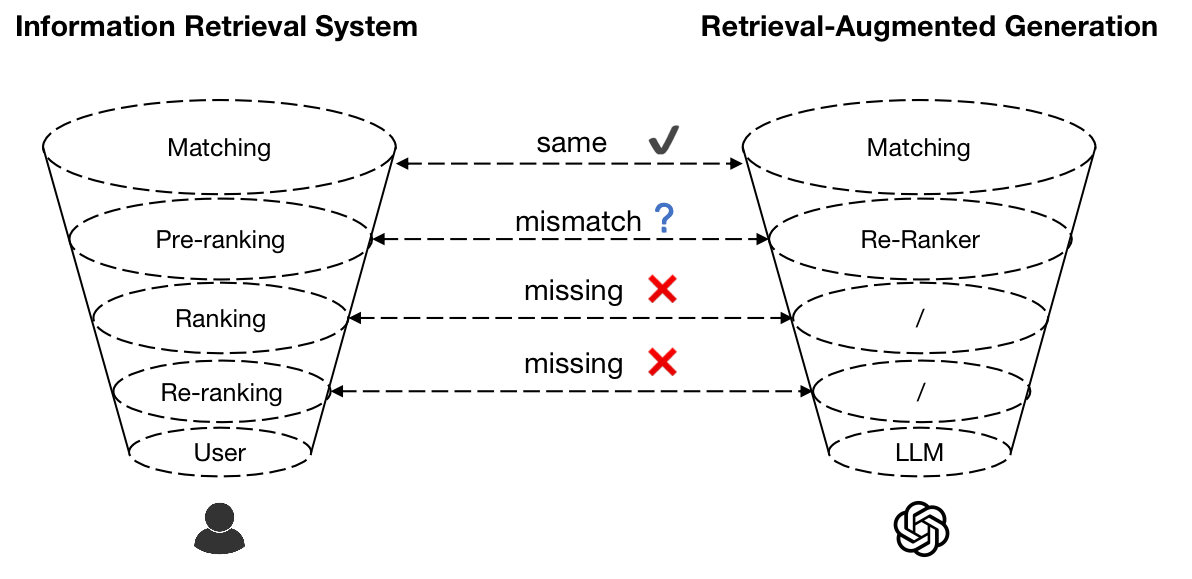}
\caption{Comparison between traditional Information Retrieval systems (such as Search Engines and Recommendation Systems) and Retrieval-Augmented Generation systems. The illustration highlights the differences in ranking processes, with RAG systems lacking a distinct ranking phase for alignment.}
\label{fig:ir_vs_rag}
\end{figure}


\begin{figure*}[h]
  \centering
  \includegraphics[width=\linewidth]{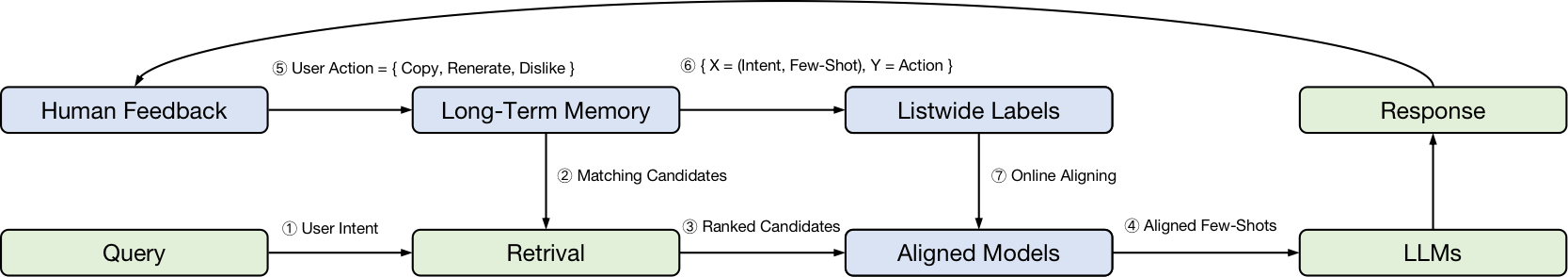}
    \caption{Pipeline of Pistis-RAG. The light blue sections represent the offline alignment process, while the light green sections correspond to the online request handling process. By collecting user copy, regeneration, and dislike actions, this feedback is integrated into the system's long-term memory as training data for Pistis-RAG alignment. This allows the RAG system to optimize search results during online user requests, aligning the retrieved content more closely with the LLM and user preferences, thereby generating content that better meets user expectations.}
    \label{fig:pipline}
\end{figure*}

We rethink RAG as a \textbf{content-centric} process, focusing on interactions from content to content. Human feedback provides end-to-end supervision, offering direct insights into content quality and prompting a re-evaluation of the RAG problem. This emphasizes the need to optimize content ranking and retrieval mechanisms to better meet human needs and enhance model performance. As illustrated in Figure \ref{fig:ir_vs_rag}, existing RAG systems face significant issues from a content-centric perspective, failing to align with LLMs and human preferences, similar to challenges in traditional information retrieval (IR) systems. Additionally, existing models labeled as ``re-ranker'' are misleading, as they primarily use representation learning and function more as pre-rankers than true rankers.

In this paper, we introduce \myname{}\footnote{The Pistis-RAG framework is named after "Pistis" from Greek mythology, symbolizing trust and reliability, which reflects the framework’s objective of systematically aligning with human values.}, the first RAG framework designed to align with human feedback and maintain continuous alignment between the ranking model, human preferences, and LLM sequencing preferences. \myname{} operates in two main phases (Figure \ref{fig:pipline}): feedback alignment and online querying.

In the feedback alignment phase, human feedback is utilized through online learning to improve the sensitivity of the ranking model to both human and LLM preferences and to adapt to evolving expectations. Our training emphasizes feedback from entire response lists rather than individual documents, necessitating strategies for effective list-wide feedback integration. To validate our method, we propose an approach that simulates human feedback using public datasets \cite{hendrycks2020measuring, huang2024c}. This technique serves as an offline pre-training and validation method, particularly advantageous for cold-start scenarios (refer to Figure \ref{fig:pipline}, specifically steps 5-7) in industrial applications.

In the querying phase, \myname{} employs a genuine ranker to reorder retrieved content based on the refined ranking model. This ranker addresses content sequencing complexities by considering both semantic relevance and presentation order to the LLM, ensuring that the final output aligns with human preferences and remains coherent with the LLM’s generative capabilities.

Our contributions are summarized as follows:

\begin{itemize}
\item \textbf{RAG Optimization via Online Learning:} We propose a method that leverages online human feedback to optimize RAG systems. \myname{} operates in two phases: feedback alignment and online querying, aligning with human preferences and adapting to evolving requirements.

\item \textbf{Ranking Model Aligned with LLM Preference:} We address the gap in the ranking stage of RAG systems from a content-centric perspective and introduce the List-Wide Feedback Alignment Model. This approach, which integrates strategies for effective list-wide feedback, addresses the limitations of existing methods that focus solely on semantic similarity and neglect LLM sequencing preferences.

\item \textbf{Validation Method for Robust Performance:} We present a validation method using public datasets to simulate human feedback. This method serves as an offline validation tool for cold-start scenarios, offering a mechanism for pre-training and evaluating RAG models in diverse operational contexts. Experimental results demonstrate that \myname{} achieves a 6.06 and 7.08\% improvement compared to baseline RAG in English and Chinese Scenario, highlighting its effectiveness in addressing existing method limitations.
\end{itemize}


%% file: related.tex
\section{Related work}

\subsection{RAG Approaches and Systems}
\begin{figure}[h]
  \centering
  \includegraphics[width=\linewidth]{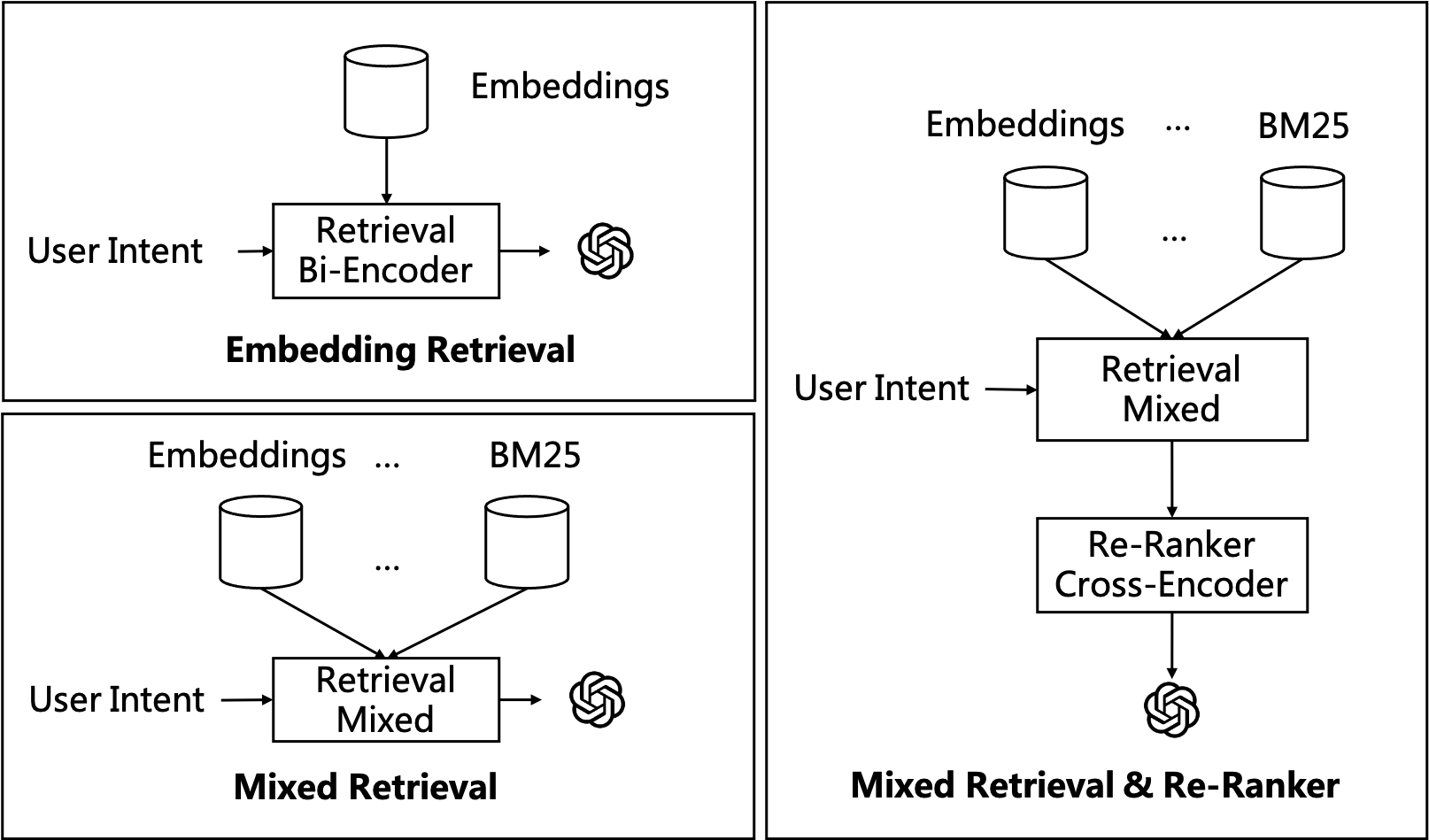}
  \caption{Brief of RAG Framework: from the initial bi-encoder encoding to mixed retrieval, and cross-encoder re-ranking.}
  \label{fig:cascade}
\end{figure}

RAG systems have evolved through several stages (see Figure \ref{fig:cascade}). Initially, representation learning techniques such as Bi-Encoders \cite{chen2024bge} and Maximum Inner-Product Search \cite{shrivastava2014improved,mussmann2016learning,douze2024faiss} were employed. These methods efficiently retrieved content with high semantic similarity. Later, additional retrieval technologies like BM25 were introduced to diversify the retrieved items, leading to the development of mixed retrieval RAG systems. However, these methods often required segmenting the original text, which could affect the balance between local semantic content and user intent during retrieval. To address this issue, re-ranker technologies, primarily Cross-Encoders, were integrated into the RAG pipeline. This enhancement improves the comparison of complete local semantic content after retrieval.

Despite these advancements, current RAG systems still face significant challenges. First, complete local semantics may not fully capture the global themes of the document or the overall memory. Recent research addressing this challenge \cite{edge2024local}. From a different angle, these requirements can be conceptualized as memory compression within intelligent agent frameworks. For example, the reflection tree proposed in \cite{park2023generative} offers a novel approach to this problem.

Moreover, optimizing semantic relevance in ranking does not necessarily ensure that LLMs produce optimal results. Studies in ICL reveal that the order of few-shot examples can significantly affect LLM performance \cite{lu2021fantastically}. Similarly, RAG tasks are sensitive to the ordering of examples \cite{yu2023augmentation}. However, the semantic similarity ranking in RAG methods often does not align with the sequences preferred by LLMs, as many LLMs have not explicitly incorporated semantic ordering in their instruction tuning \cite{dubey2024llama, achiam2023gpt, bai2023qwen}. Research addressing this issue includes \cite{ke2024bridging, yu2023augmentation,dong2024understand}. Those approaches mainly rely on offline methods and do not effectively leverage high-quality human feedback signals. In addition, there is limited research on how to model these signals and apply online learning optimization. Our work addresses these gaps by utilizing human feedback for supervision and incorporating this feedback into a two-phase online learning framework to enhance RAG performance continuously.

\subsection{LLMs and Online GenAI Systems}
The rapid emergence of online GenAI systems has escalated the demand for their online performance capabilities. Research in this domain encompasses several critical areas, which not only affect user experience but also influence the cost efficiency of related services. Firstly, enhancing the online inference efficiency of large-scale models is a primary focus, with initiatives such as \cite{kwon2023efficient} dedicated to performance optimization aimed at improving computational resource utilization. Furthermore, an evolving body of research is addressing issues of fairness in online GenAI services. This notion of fairness is defined concerning specific characteristics of GenAI applications, as illustrated by studies such as ~\cite{sheng2024fairness}. The integration of various types of online human feedback as an optimization signal in online GenAI systems is still in its early stages. WebGPT \cite{nakano2021webgpt} and RLHF \cite{ouyang2022training} are notable methods that efficiently leverage human feedback. We aim for our research to contribute to further investigation and discussion on this topic, to improve how GenAI systems align with human values.

%% file: approach.tex
\section{\myname{} Approach \& Pipeline}

We now unpack the high-level data flow of our approach and pipeline in Figure \ref{fig:pipline}, describing key design parameters, techniques, and implementation details for each step. Since the aligning phase serves as a prerequisite for the query phase, we follow this order in the following section.

\subsection{Human Feedback → Long-Term Memory}
A pivotal component of our methodology is the integration of human feedback into long-term memory. This feedback is crucial to align the system with user preferences and serves as prospective content for future quick response. 



In online data collection, two primary challenges arise. First, the rapid growth in data volume increases storage costs and reduces retrieval efficiency. Second, excessive content homogeneity affects the diversity of information after retrieval and truncation. We address these challenges using de-duplication and filtering techniques, which help us manage and optimize online datasets, improving storage and retrieval efficiency while preserving data quality and diversity.

\paragraph{De-duplication}
We apply the MinHash algorithm \cite{broder1997resemblance} for de-duplication. MinHash approximates set similarity and identifies duplicate documents within the dataset, which reduces storage requirements and improves retrieval efficiency.

\paragraph{Filtering}
We use Kullback-Leibler (KL) divergence based on token distribution for filtering. KL divergence measures the difference between probability distributions, allowing us to filter out documents with outlier tokens. This helps reduce noise and improve the quality of the data used for retrieval.

\paragraph{Memory Structure}
By effectively indexing and retrieving this feedback, the system can gradually adapt to user preferences and enhance its performance over time. The dataset is represented as:

\[
\mathcal{D} = \{ (P_I, X_I, Y_I, F_I) \}_{I=1}^N
\]

where:

\begin{itemize}
    \item \( P_I \) is the user's intent,
    \item \( X_I = \{ x_{I1}, x_{I2}, \ldots, x_{Ik} \} \) is an ordered list of \( k \) few-shot examples,
    \item \( Y_I \) is the generated output,
    \item \( F_I \) is the list-level label indicating user feedback for the \( I \)-th list \( X_I \).
\end{itemize}



\subsection{Long-Term Memory → Aligned RAG System} \label{subsec:aligned_ranker}
In this phase, we integrate long-term user feedback into the RAG system to align it with LLMs and user preferences. By incorporating user feedback, we optimize the sequence presented to LLMs, thereby adjusting the item order according to LLM's preferences for content ranking. Feedback is collected through user interactions and feedback mechanisms, using each few-shot prompting list \( X_I \) and its corresponding prompt \( P_I \) as features. The feedback label \( F_I \) indicates the user's assessment and is categorized as follows:

\begin{itemize}
  \item \( F_I = \text{Copy} \): Content-copying feedback, indicating that the content was accepted.
  \item \( F_I = \text{Regenerate} \): Content-regeneration feedback, suggesting that the content should be revised or regenerated.
  \item \( F_I = \text{Dislike} \): Content-disliking feedback, denoting that the content was rejected.
\end{itemize}

These labels represent both explicit user preferences and implicit preferences of the LLM. To ensure ongoing alignment between the model and user feedback as well as LLM preferences, it is essential to employ both offline and online learning methodologies. Consequently, this consideration should guide the selection of methods. We leverage \textit{RankFormer} \cite{buyl2023rankformer} to supervise end-to-end listwide feedback signals in an extension of the listwise Transformer architecture. This approach maintains robust performance even after we distill the model to Gradient Boosted Decision Trees \cite{friedman2001greedy}.

\subsection{Query → Memory → Aligned Few-Shot Examples}
We process queries to retrieve relevant data from long-term memory, resulting in a refined list of candidate examples. We then use the aligned RAG ranker, which supports online learning, to reorder these examples based on alignment with model preferences and real-time user feedback. This capability enables the model to adapt dynamically to streaming data and continuously align with user preferences. Consequently, the sequence of few-shot examples remains well-suited to generate accurate and contextually appropriate responses.

\subsection{Aligned Few-Shot Examples → Answer}
The final step uses the aligned few-shot examples to generate the final response. We incorporate user feedback collected during this phase to refine and enhance the system’s performance. This feedback loop ensures that the generated responses meet user expectations and continuously improve alignment with human preferences.

%% file: evaluation.tex
\section{Evaluation}


Our experiments reveal the impact of our aligned ranking model on both the latency and accuracy of the RAG system.

\subsection{Datasets}
To mitigate potential biases from open-domain question-answering evaluation methods, such as N-Gram or Exact Match, which could affect the validity of our experimental results, we use datasets (refer to Table \ref{tab:dataset_overview}) consisting of multiple-choice questions with objective answers. This approach allows us to ensure a more reliable assessment by focusing on questions with clearly defined correct responses.

\begin{table}[h!]
\centering
\begin{tabular}{lcc}
\toprule
\textbf{Dataset} & \textbf{Number of Subjects} & \textbf{Data Size} \\
\midrule
MMLU & 57 & 15,858 \\
C-EVAL & 52 & 13,948 \\
\bottomrule
\end{tabular}
\caption{Overview of datasets: the dataset name, number of subjects, and the size of the dataset.}
\label{tab:dataset_overview}
\end{table}


\paragraph{MMLU (English)} The MMLU dataset is a key resource for researchers developing NLP systems and has been used to evaluate and improve various advanced NLP systems. In our experiments with the MMLU dataset, we use 1,540 validation samples for retrieval-augmented content and 14,079 test samples as queries to evaluate the RAG system.


\paragraph{C-EVAL (Chinese)} The C-Eval dataset is a key resource for researchers developing Chinese NLP systems. It is designed to assess a framework's ability to understand and generate coherent responses across various subjects particularly suited for evaluating the \myname{} framework's performance in handling complex Chinese queries. For our evaluation, we use 260 development samples for retrieval-augmented content and 14,079 validation samples as queries to assess the RAG system. We select the development and validation sets because the test set does not provide correct answers.



\subsection{Generators}
In our practical implementation, RAG is primarily utilized for online scenarios. Consequently, we employ a high-performance LLM with approximately 10 billion parameters, selected for its cost-efficiency, practicality, and particularly its favorable Time To First Token (TTFT), as the generator for simulation, alignment, and evaluation. Context Window is also an important metric which is a limitation of RAG candidate size (refer to Table \ref{tab:model_comparison}).

\begin{table}[h!]
\centering
\begin{tabular}{lcc}
\toprule
\textbf{Model} & \textbf{Training Data} & \textbf{Context Window} \\
\midrule
LLaMA-2 & 2 trillion tokens & 4K tokens \\  
Qwen-2 & 4-7 trillion tokens & 128K tokens \\
\bottomrule
\end{tabular}
\caption{Comparison of LLaMA and Qwen models on key metrics: training data and context window size.}
\label{tab:model_comparison}
\end{table}

\paragraph{Llama-2} The Llama-2 model \cite{touvron2023llama} is representative of open source LLM, excelling in natural language understanding and generation, providing coherent responses across various NLP tasks.

\paragraph{Qwen-2} To identify a representative LLM for Chinese language tasks, we select Qwen-2. The tech report \cite{yang2024qwen2} shows that Qwen2-7B outperforms most other models across various datasets and excels in Chinese language tasks.

\begin{figure*}[h]
  \centering
  \includegraphics[width=.95\linewidth]{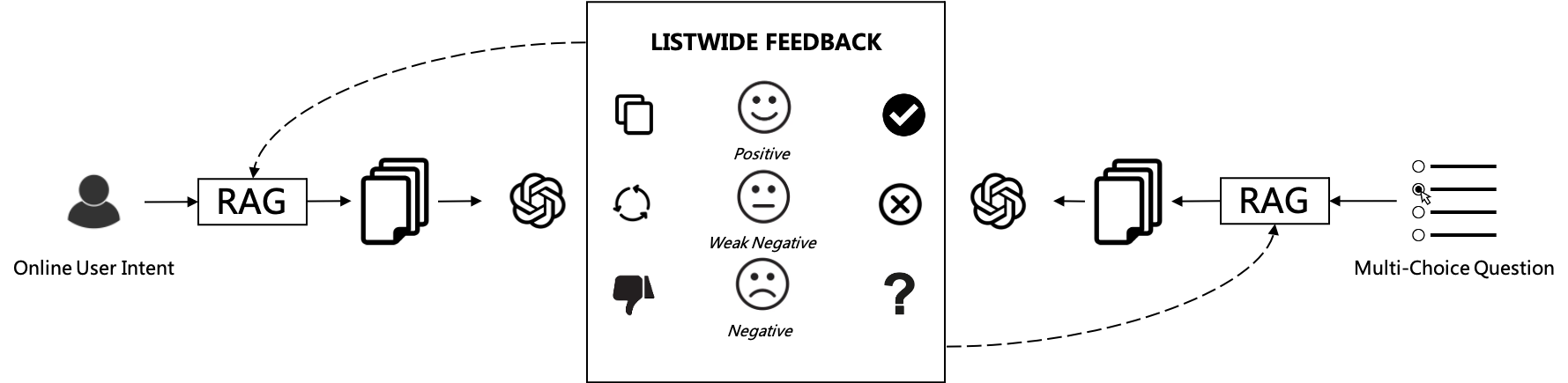}
    \caption{Simulating Feedback: Comparison of online human feedback versus simulated feedback. Correct answers (\(Y_{\text{correct}}\)) are represented by text copying, incorrect answers (\(Y_{\text{incorrect}}\)) reflect regeneration, and no answers (\(Y_{\text{no\_answer}}\)) indicate negative user feedback.}
    \label{fig:feedback_simulation}
\end{figure*}
\subsection{Simulating Feedback}\label{subsec:pipline}

\begin{table*}[h!]
    \centering
    \caption{Comparison of RAG End-to-End Performance Metrics (F1-Score) for English and Chinese Languages}
    \begin{tabularx}{.85\textwidth}{X|cc}
        \toprule
        \textbf{Metric \textbackslash  Dataset} & \textbf{English (MMLU)} & \textbf{Chinese (C-EVAL)} \\
        \midrule
        \textbf{Total Number of Indexes} & 1,540 (validation) & 260 (development) \\
        \textbf{Total Number of Queries} & 14,079 (test) & 1,346 (validation) \\
        \textbf{Generator Models} & LLaMA-2-13B-Chat (5-shot) & Qwen2-7B-Instruct (5-shot) \\
        \midrule
        \textbf{Naïve RAG Accuracy} & 49.6027 & 77.8698 \\
        \textbf{Pistis-RAG Accuracy} & \textbf{52.6088} & \textbf{83.3901} \\
        \textbf{Accuracy Improvement ($\Delta$)} & \textcolor{red}{\textbf{+6.0603\%}} & \textcolor{red}{\textbf{+7.0891\%}} \\
        \textbf{Latency Impact ($\Delta$)} & +16.3109\% & +17.7204\% \\
        \bottomrule
    \end{tabularx}
    \label{tab:result}
\end{table*}

Online platforms, such as ChatGPT, continuously enhance their performance through user feedback mechanisms such as copying, regenerating, and disliking responses. Our simulation process aims to align our system with this prevalent scenario.

\subsubsection{Collecting Few-shot Examples}

Within our system architecture, we seamlessly integrate highly rated user feedback into our contextual learning example database for the utilization of RAG. Specifically, we incorporate precise question-answer pairs sourced from public datasets as supplementary ICL examples. To ensure impartial evaluation, we filter out answers identical to questions during retrieval (though in the live online system, such cases are directly cached for future retrieval). This meticulous approach establishes a simulated example of learning from an external knowledge base, enriching the system's performance through augmented retrieval data.

\subsubsection{Simulating User Behavior}\label{subsubsec:sim_user}

\begin{figure}
    \centering
    \includegraphics[width=\columnwidth]{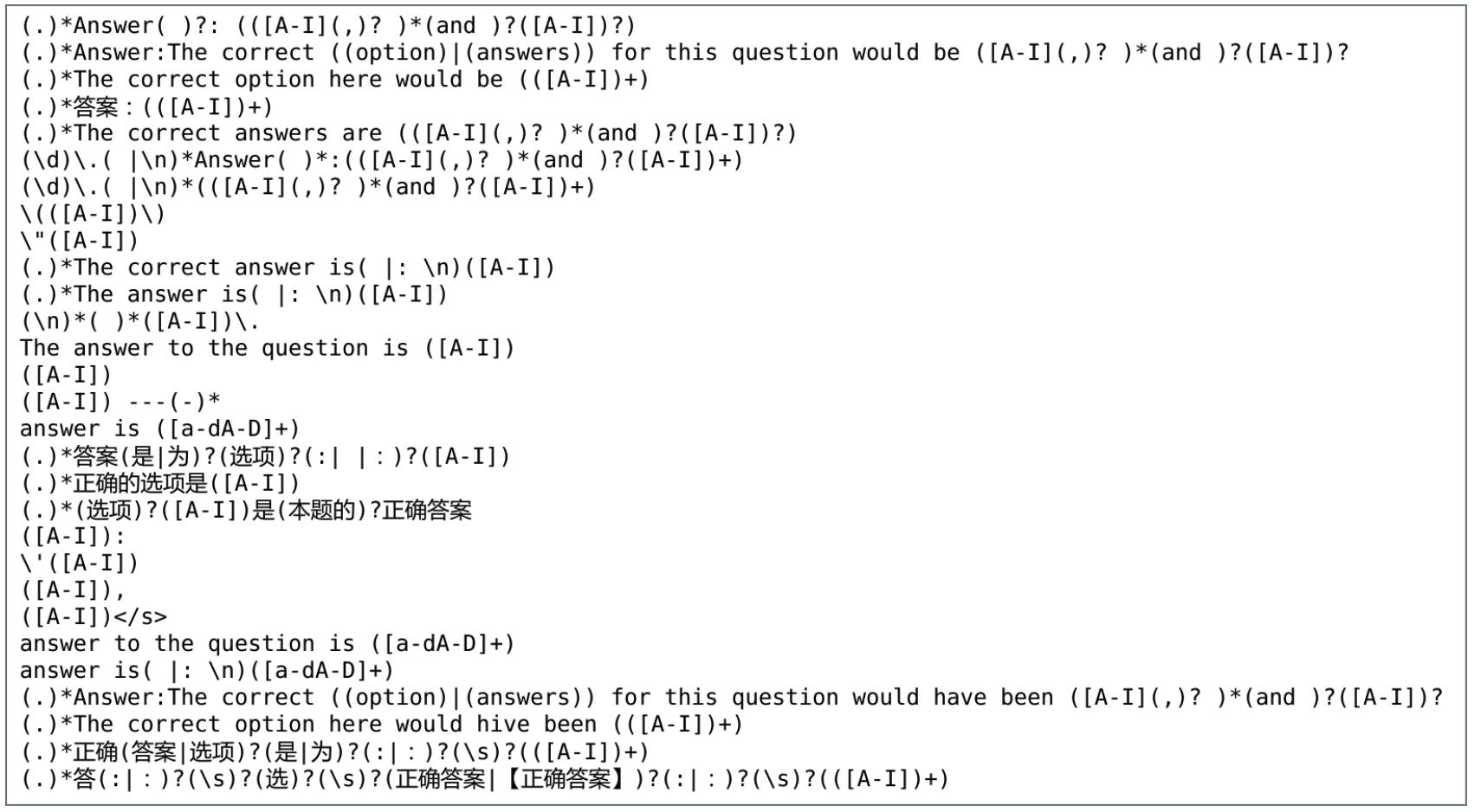}
    \caption[Pistis-RAG Framework]{RegEx patterns used in RegEx-Based Extraction for extracting feedback.}
    \label{fig:regex}
\end{figure}

Figure \ref{fig:feedback_simulation} shows the user behavior simulation process. We refine user behavior data from open-source datasets to simulate user feedback through the following steps:

\begin{enumerate}
    \item \textbf{Leveraging Retrieval-Augmented Generation:} Our system employs RAG to augment response generation by retrieving relevant information from an indexed dataset in Table \ref{tab:result} before generating an answer. This process aims to establish the correlation between RAG ranking methods and generated responses.
    
    \item \textbf{Extracting Information with Regular Expressions:} Following the generation stage, we employ regular expressions to extract specific information \(X\) from the generated texts \(Y\). This operation is represented as \(P(Y) = X\), where \(P\) represents the regex parsing function in Figure \ref{fig:regex}.
    
    \item \textbf{Assigning Labels Based on Correctness:} The final step involves assigning labels to the generated outputs based on their accuracy relative to the expected answers. We define the labeling function \(L(y)\) for the generated output \(y\) in comparison to the expected answer.
\end{enumerate}

This feedback loop is fundamental to continuously enhancing the model's accuracy and aligning it with user preferences. Here's the breakdown of the labels:
\begin{itemize}
    \item \textbf{Correct (\(L(y) = Positive\)):} Indicates that the generated output \(y\) matches the set of correct answers \(Y_{\text{correct}}\).
    \item \textbf{Incorrect (\(L(y) = Even\)):} Signals that the generated output \(y\) falls within the set of incorrect answers \(Y_{\text{incorrect}}\).
    \item \textbf{No Answer (\(L(y) = Negative\)):} Represents outputs lacking an answer and belonging to the set of nonresponses \(Y_{\text{no\_answer}}\).
\end{itemize}

This formalization illustrates a continuous learning system in which feedback from real user interactions refines and enhances the model, aligning its output more closely with user expectations and real-world applications. Specifically, the correct answers (\(Y_{\text{correct}}\)) resemble text copying, the incorrect answers (\(Y_{\text{incorrect}}\)) resemble regeneration, and the lack of answers (\(Y_{\text{no\_answer}}\)) correspond to negative user feedback.

\subsubsection{Traning Set Sampling}
In our methodology, the theoretical number of combinations for selecting \( k \)-shot examples from \( n \) examples is represented by \( \binom{n}{k} \). Here, \( n \) is specific to the retrieval-augmented dataset in use, such as the MMLU training set or the C-EVAL development set. Evaluating all possible combinations is computationally prohibitive and poses a risk of overfitting the model.

To address these issues, we use the Monte Carlo method. This technique involves randomly sampling examples from the dataset to create few-shot examples, which significantly reduces computational costs and helps prevent overfitting. Subsequently, we simulate feedback, requiring approximately 2 GPU years of computation using A800 GPUs. We then apply stratified sampling to average the labeled samples, resulting in a final supervised dataset of 10 million 5-shot examples with feedback labels, which is used to train the aligned ranking model.

\subsection{Experimental Setup}
Our experimental setup is meticulously designed to optimize the performance of our models in both retrieval and response generation. Below are the key components of our setup:

\paragraph{Indexing} To assess the efficacy of our proposed method, we conducted experiments utilizing the public dataset. Initially, we constructed an index for the MMLU (C-Eval) training (development) set. 

\paragraph{Matching} We employed BEG-M3 to retrieve the top 10 candidate using Milvus\cite{wang2021milvus}. BEG-M3 functions as a dense vector representation model for similarity search, adept at efficiently retrieving analogous vectors from an extensive vector repository. Milvus serves as a vector database, designed to store and manage vector representations.

\paragraph{Pre-Ranking} We apply BEG-reranker-larger to pre-rank these candidates and select the top five. We use these top candidates to generate 5-shot results and establish the \textbf{naïve RAG baseline} for evaluation. We also use these candidates as input for further analysis in the ranking stage.

\paragraph{Ranking} 


We use the settings mentioned in \textit{Long-Term Memory → Aligned RAG System}\ref{subsec:aligned_ranker} and the feedback simulation methods described in \textit{Simulating Feedback}\ref{subsubsec:sim_user} to generate data and train an end-to-end listwide label learning-to-rank model. 

This model reorders the top candidates selected after pre-ranking selection to align with the LLM's sequential preferences and the human value preferences reflected in the task's ground truth from public datasets.

\paragraph{Few-Shot Prompt} Few-shot prompts leverage the selected set of example instances. These prompts are fed into a conversational model, which then generates textual outputs based on the data and patterns presented in the examples.

\begin{quote}\begin{scriptsize}\begin{verbatim}

How many numbers are in the list 25, 26, ..., 100?
(A) 75 (B) 76 (C) 22 (D) 23
Answer: B
...
[additional exemplar instances]

If 4 daps = 7 yaps, and 5 yaps = 3 baps,
how many daps equal 42 baps?
(A) 28 (B) 21 (C) 40 (D) 30

Please provide your answer as a single character
(e.g., 'A', 'B', 'C', 'D', etc.):

\end{verbatim}\end{scriptsize}\end{quote}

\begin{figure}
    \centering
    \includegraphics[width=\columnwidth]{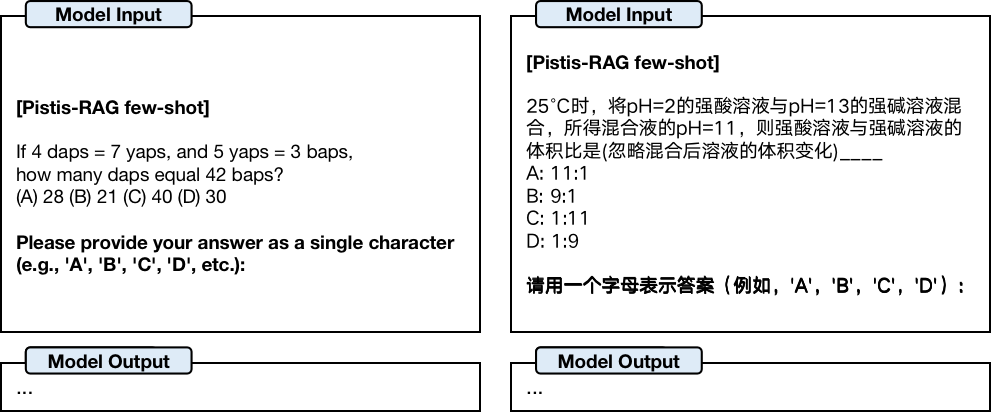}
    \caption[Pistis-RAG Framework]{Comparison of Chinese and English prompt templates.}
    \label{fig:prompt}
\end{figure}

This example is based on the prompt template in Figure \ref{fig:prompt}. We then apply the method described in \textit{Simulating User Behavior}\ref{subsubsec:sim_user} to extract answers from the content generated by this few-shot prompt and to produce feedback labels.


\subsection{Experimental Results}\label{subsec:results}

\paragraph{Accuracy} The results demonstrate a notable improvement in accuracy for the Pistis-RAG framework across both English and Chinese datasets. Specifically, the alignment model shows a significant enhancement in the accuracy metric, with an increase of 6.06\% in MMLU (English) and 7.08\% in C-EVAL (Chinese), indicating the effectiveness of our approach in improving model alignment with user preferences.

\paragraph{Latency}

The results show that latency increased by 16.31\% for MMLU (English) and 17.72\% for C-EVAL (Chinese), reflecting the computational cost of alignment. The subtle differences in end-to-end latency between the datasets can be attributed to two main factors. First, variations in long-term memory storage can cause minor delays during retrieval. Second, fluctuations in experimental resources can lead to slight discrepancies in latency across repeated tests, although these variations generally do not significantly impact overall consistency. Additionally, differences in generation performance between Qwen-2 and Llama-2 result in varying time ratios spent by the generator.

While the latency variation data is provided in the context of optimizing RAG's end-to-end accuracy, it highlights that increased latency is an expected outcome. This reflects an optimization trade-off rather than a simple gain-loss scenario. For more details, refer to \textit{Discussion}\ref{sec:discussion} in the next section.

%% file: discussion.tex
\section{Discussion} \label{sec:discussion}

In practice, we treat the truncation lengths for matching, pre-ranking, and ranking as global parameters of the RAG system. By optimizing these parameter combinations, users can select the parameters that best balance the business needs for latency and accuracy. We enable users to choose the truncation length that fits their requirements, as we do in our industrial scenarios, allowing them to improve accuracy without increasing latency, or even reducing it (though not to the extreme accuracy improvement).

In RAG systems, accuracy and latency are critical performance metrics. Optimizing these metrics, particularly when system latency approaches its maximum allowable threshold, presents a complex challenge. This section examines downgrade strategies, the effects of introducing the ranking phase, the formalization of optimization problems, and parameter adjustments in online services.

\begin{figure}[h]
  \centering
  \includegraphics[width=.9\linewidth]{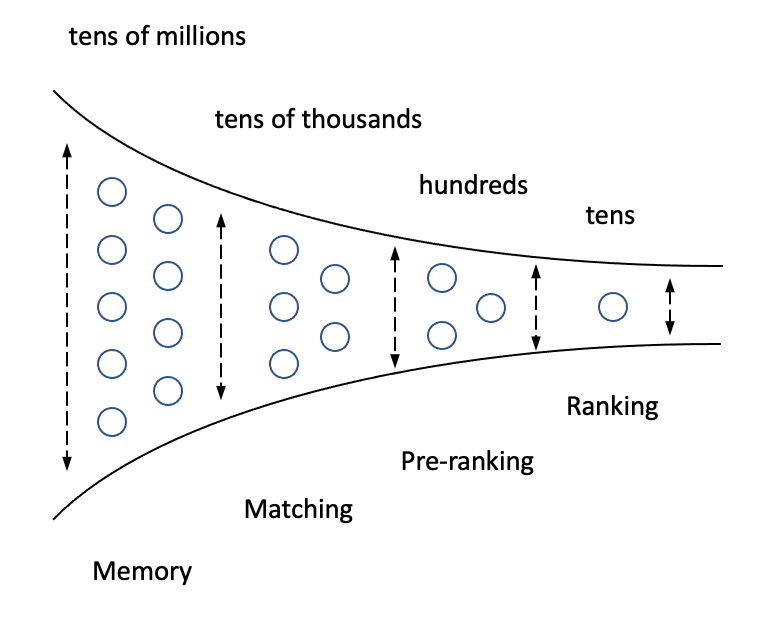}
  \caption{An Illustration of a Content-Centric Cascading System (\myname{}): This diagram highlights the process where content retrieval and generation are from a content-centric view, utilizing external data sources.}
  \label{fig:content-centric}
\end{figure}

\paragraph{Downgrade Strategies}

Downgrade strategies become essential when system latency nears the maximum allowable threshold. This threshold, denoted as $L_{th}$, represents the maximum acceptable latency for the system. When system latency $L$ reaches or exceeds $L_{th}$, it is necessary to reduce the service level to maintain system stability. When latency is below $L_{th}$, service quality can be enhanced. This approach helps maintain system stability and responsiveness under varying load conditions.

\paragraph{Impact of Introducing the Ranking Phase}

The introduction of the ranking phase aims to improve system accuracy. However, this may result in an increase in system latency. Experimental results \ref{subsec:results} showing changes in accuracy and latency illustrate this impact. Although the data may indicate that the ranking phase could increase latency, this is part of an optimization problem rather than a simple trade-off.

In RAG systems, truncation parameters for matching, pre-ranking, and ranking are global settings. Adjusting these parameters can balance the requirements for latency and accuracy. Selecting the appropriate truncation parameters based on specific business needs is recommended. This tuning can improve accuracy while keeping latency stable, though it may not achieve the optimal results. This reflects the practical application of a content-centric approach.

\paragraph{Formalization of the Optimization Problem}

Formally, truncation parameters for matching, pre-ranking, and ranking are variables. Fixing any two parameters and adjusting the third parameter typically results in latency being negatively correlated with these parameters, while accuracy is positively correlated. Consequently, optimizing accuracy under a given latency constraint constitutes an optimization problem. The mathematical model can be expressed as:
\[
\text{Maximize } \text{Acc} \text{ subject to } \text{Lat} \leq L_{th}
\]
where $\text{Acc}$ represents accuracy, $\text{Lat}$ represents latency, and $L_{th}$ is the system’s maximum allowable latency.

\paragraph{Parameter Adjustment in Online Services}
In online service operations, system latency may vary. To manage latency effectively, particularly when it approaches the maximum allowable threshold $L_{th}$, it is advisable to adjust one or more of the truncation parameters. This adjustment helps optimize latency while maintaining system accuracy, ensuring stability under high-load conditions. Specifically, if the system latency $L$ reaches or exceeds $L_{th}$, reducing any of the parameters $(p_m, p_{pr}, p_r)$ in Figure \ref{fig:content-centric} can effectively control latency. Detailed guidance on optimal parameter settings for industry applications is available in \textit{Appendix}\ref{sec:appendix}.



%% file: conclusion.tex
\section{Conclusion}

We introduce a content-centric RAG approach that integrates list-wide label learning to rank and online learning methodologies to ensure precise alignment with human feedback. The framework operates in two primary phases: feedback alignment and online querying. Multilingual evaluations conducted on two extensively used public datasets demonstrate that our method significantly surpasses the performance of the traditional RAG baseline in both English and Chinese contexts.

These evaluation results establish a basis for an in-depth analysis of the latency-accuracy trade-off. We propose a methodology for exploring this balance by adjusting truncation parameters, and formalizing it as an optimization problem specific to online systems. This approach enables systematic assessment and optimization of parameters to maximize system performance while adhering to constraints on maximum response latency.

In summary, our method not only effectively aligns the RAG system with human feedback but also provides a highly efficient online serving solution.